# Microwave Integrated Circuits Design with Relational Induction Neural Network

Jie Liu[1], Zhi-Xi Chen[1,*], Wen-Hui Dong[2], Xiao Wang[3], Jia Shi[1], Hong-Liang Teng[1], Xi-Wang Dai[4], Stephen S.-T. Yau[2], Chang-Hong Liang[5], Ping-Fa Feng[3].

1, Shijiazhuang Chuangtian Electronic Technology Co., Ltd, Shijiazhuang, 050091, China

2, Department of Mathematical Sciences, Tsinghua University, Beijing, 100084, China

3, Department of Mechanical Engineering Tsinghua University, Beijing, 100084, China

4, Department of Electronics and Information, Hangzhou Dianzi University, Hangzhou, 310018, China

5, Department of Electronics and Information, Xidian University, Xian, 710071, China

*Correspondence should be addressed to Zhi-Xi Chen (ctkj_rf@163.com)

**Abstract**: The automation design of microwave integrated circuits (MWIC) has long been viewed as a fundamental challenge for artificial intelligence owing to its larger solution space and structural complexity than Go. Here, we developed a novel artificial agent, termed Relational Induction Neural Network, that can lead to an automotive design of MWIC and avoid brute-force computing to examine every possible solution, which is a significant breakthrough in the field of electronics. Through the experiments on microwave transmission line circuit, filter circuit and antenna circuit design tasks, strongly competitive results are obtained respectively. Compared with the traditional reinforcement learning method, the learning curve shows that the proposed architecture is able to quickly converge to the pre-designed MWIC model and the convergence rate is up to four orders of magnitude. This is the first study which has been shown that an agent through training or learning to automatically induct the relationship between MWIC's structures without incorporating any of the additional prior knowledge. Notably, the relationship can be explained in terms of the MWIC theory and electromagnetic field distribution. Our work bridges the divide between artificial intelligence and MWIC and can extend to mechanical wave, mechanics and other related fields.

## Introduction

Microwave integrated circuits are sparked by the collision of human wisdom, experience and intuition. Engineers use computer-aided design tools to analyze and solve MWIC problems and then try to find the best solution[1-3]. This process is extremely tedious, boring and inefficient. Limited by human physiological structure, engineers can hardly find the optimal solution of large-scale MWIC. How to make engineers break through these bottlenecks is very useful.

At present, all the researchers abstract MWIC parameters artificially, and then optimize these parameters with machine learning technology[4-19]. However, this approach has two serious drawbacks: first, it is a time-consuming and laborious work, and the abstracted parameters may not be accurate enough to represent some important features of the circuits; second, it will greatly limit the imagination and exploration space of an agent, resulting in who is often difficult to exceed the human level.

In recent years, the field of artificial intelligence[20,21] (AI) has enjoyed many successes in data mining[22], computer vision[23], natural language processing[24] and other fields of application. As a sub-field of AI, reinforcement learning[25-27] (RL) based on deep neural networks has gradually shifted from pure academic research to application, such as classic video-games[28], board games[29], neural machine translation[30] and drug design[31]. However, it remains empty on how to combine AI with MWIC design. Due to the complex structure and a huge solution space of MWIC design, traditional RL algorithms need to rely on mass data to learn the design decision-making process, which make them difficult to converge quickly in time. Therefore, we create an architecture termed RINN (Relational Induction Neural Network) whose structure efficiently learn the rules of MWIC data, achieving the goal of designing arbitrary complex MWIC. More specifically, the MWIC shape is defined as a set of parameterized mesh, and when each mesh changes, the simulation result is calculated by standard CAE packages such as ADS or Ansys EM. Then, the clustering algorithm of RINN is used to cluster the changes of these simulation results.

The main contributions of this paper are summarized as follows. First, to our knowledge, this is the first work that tries to explore MWIC design by training agents using deep RL methods without relying on human experience, fulfilling the gap in this respect. Second, clustering algorithm is used to reduce the solution space of MWIC design, which has much more powerful unsupervised learning ability and ensures that the RINN architecture has better stability and faster convergence speed. Third, several comprehensive studies, which are conducted on different aspects of an automotive design on microwave transmission line circuit, filter circuit and antenna circuit, have been successfully demonstrated that how to: 1) train RINN as an agent for designing MWIC; 2) integrate between MWIC design and machine learning. This method can also be used to train agents in other fields such as mechanics, pointing out a direction for future automated design.

# Results

## RINN architecture

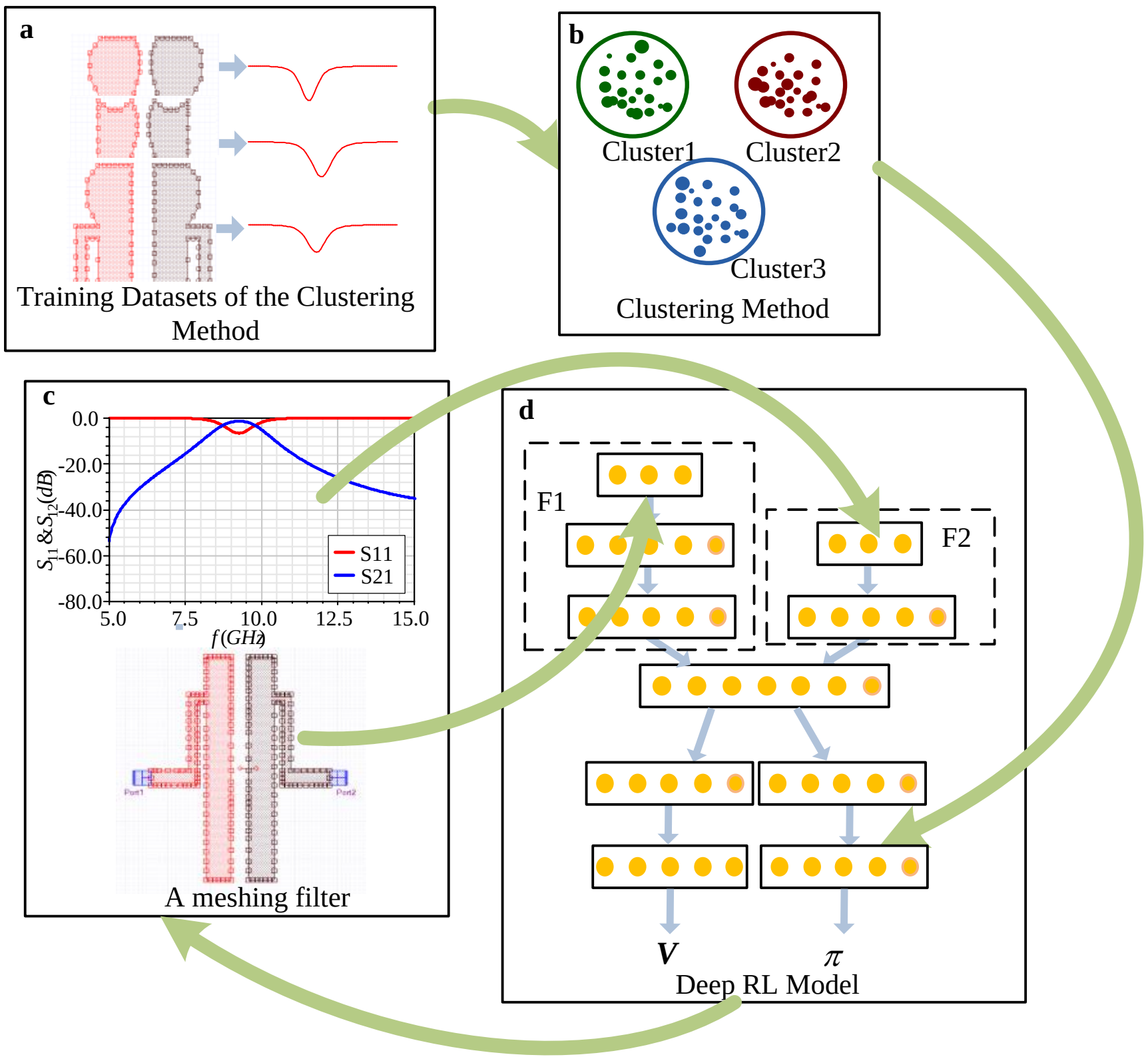

**Figure 1 | RINN architecture. a,** Datasets of clustering algorithm, i.e. S parameter change matrix of meshed model. **b,** Clustering algorithm. **c,** Meshed model and S parameter matrix to train the Deep RL model. **d,** Deep RL model with **c** as input, probability vector $\pi$ of design action for MWIC model and value scalar $V$ as output.

The RINN architecture (summarized in Fig.1) for MWIC design includes two parts: the clustering algorithm (Fig.1b) and the deep RL neural network model (Fig.1d). In this architecture, the clustering algorithm is used to cluster the design actions of meshed MWIC, i.e. many design actions of MWIC are clustered into several typical action clusters, where it plays a role of an experienced MWIC engineer to design the parameterized MWIC model; whereas the deep RL neural network model predicts the design action of the current MWIC model based the clustering algorithm results. The deep RL neural network model is addressed with the Asynchronous Advantage Actor-Critic (A3C) algorithm[32] that relies on learning both a policy and value function. Typical action clusters are used as the policy networks output action to predict the design actions of current integrated circuit models, and then the design actions are evaluated by value networks to find out the optimal policy so that the algorithm proposed in this paper can achieve the technical efficiency of automatic design. These details can also be found in the Methods.

**Filters design based RINN**

**Table 1** Four design tasks for filters

| Task | Passband /GHz | Center frequency /GHz | Insertion loss /dB | Reflection loss /dB | Filter Size L/mm*W/mm |
|---|---|---|---|---|---|
| 1 | 8.9~9.7 | 9.3 | >-0.5 | < -20 | -- |
| 2 | 11.1~11.9 | 11.5 | >-0.5 | < -20 | -- |
| 3 | 7.3~7.8 | 7.55 | >-0.5 | < -20 | -- |
| 4 | 6.7~7.2 | 6.95 | >-0.5 | < -20 | 5*5 |

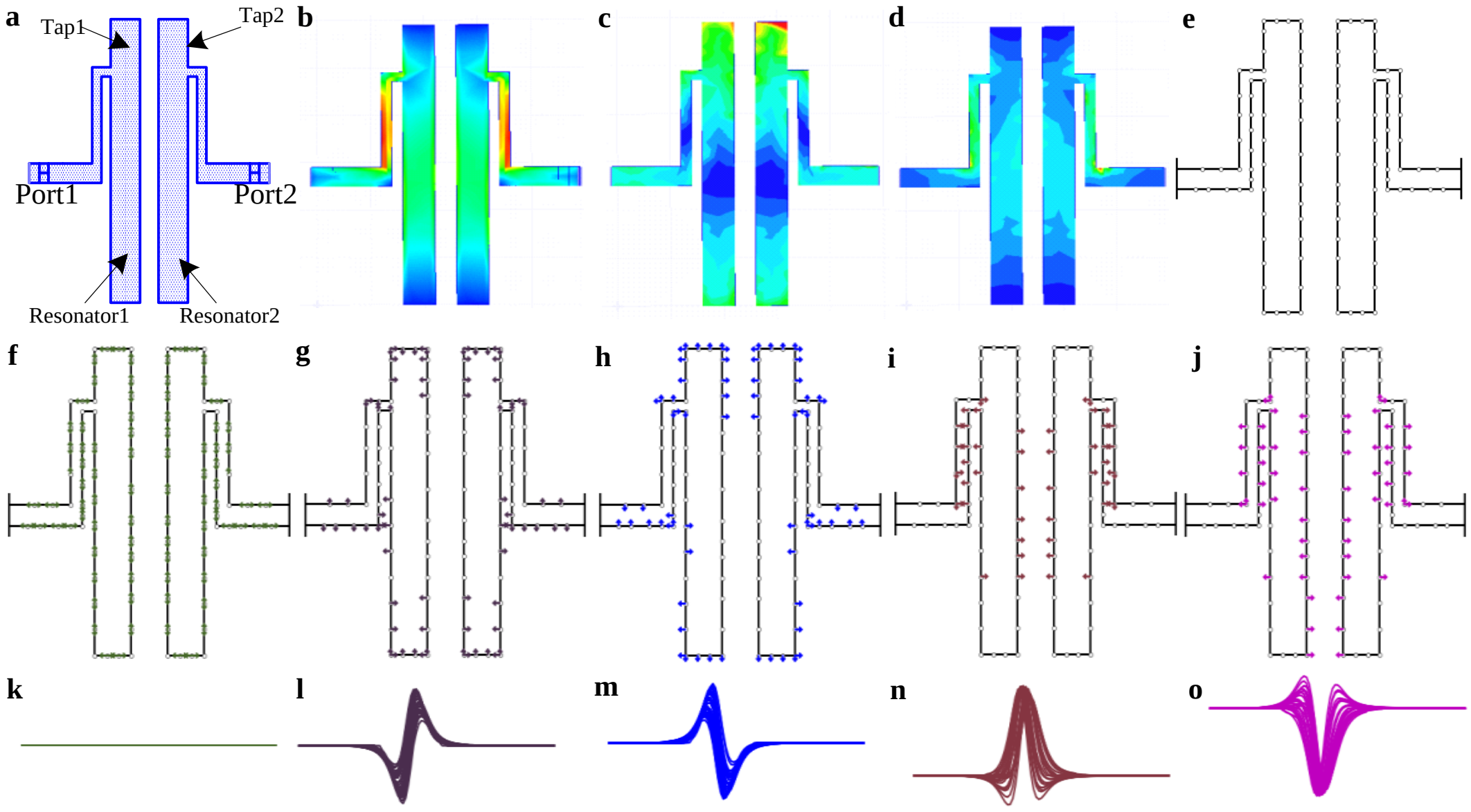

**Figure 2 | Clustering visualization results of filters. a,** Filter model. **b-d**, The surface current density distribution, electric field distribution and magnetic field distribution of the designed filter model. **e**, Meshed model. **f-j,** Visualization results of typical action clusters on the designed filter. **k-o,** Differential S11 curves of the typical action clusters.

To investigate the capacity of the RINN to perform MWIC design, we considered four design tasks, more precisely, four filters were designed whose center frequencies are 9.3GHz, 11.5GHz, 7.55GHz and 6.95GHz respectively, but the length and width of the fourth filter were limited to 5mm*5mm. The specific design tasks are as shown in Table 1. The specific design scheme was as follows: we first designed a parallel coupled Chebyshev bandpass filter as shown in Fig.2a, whose substrate was gallium arsenide material, dielectric constant was 12.9, conductor was made of gold, the resonator was made of half-wavelength resonator with open circuit at both ends, and the input-output coupling was made of tap line. Furthermore, the shape of the filter was divided by meshes (see Methods) as shown in Fig.2e. We then obtained the training datasets of clustering algorithm from the differential S parameter matrix. The differential S parameter matrix was the difference between the original S parameter matrix without actions and the S parameter matrix by performing actions of 0.05 mm, 0.1 mm and 0.15 mm on the mesh vertices shown in Fig.2e. Finally, we clustered the training set and utilized the subsequent cluster assignments as typical action clusters of MWIC design.

Fig.2 shows the visualization results of clustering. By analyzing the results based on the filter theory, we find that the design actions of the filters are abstracted into four clusters: the resonant frequency increases (Fig.2g), the resonant frequency decreases (Fig.2h), the coupling ratio increases (Fig. 2i) and the coupling ratio decreases (Fig.2j). Further analysis of the results from the electromagnetic field distribution of the filter, the surface current density distribution (Fig.2b) shows that the coupling signal is primarily concentrated at the middle of the resonators, the electric field (Fig.2c) of the filter is mainly concentrated at the open end of the resonators, and the magnetic field (Fig. 2d) is mainly concentrated at the middle of the resonators, whereas the clustering results of Fig.2(g,h) and Fig.2(i,j) show that the clustering algorithm also basically aggregates the electric field at the open end of the resonators and the magnetic field at the middle of the resonators, respectively. Therefore, the analysis of filter theory and filter electromagnetic field distribution confirm that the proposed RINN architecture can effectively extract the typical action clusters of MWIC model.

The deep RL model was set with clustering results whose sizes of policy logits is 4 (excluding the action cluster that had little effect on the filter). The action directions of the agent corresponded to Fig.2(g-j). The agent learned from scratch how to design MWIC models without the premise of informing design rules. Hopefully, we found that agents actually have learned behaviors similar to engineers' through observing the behavior of agents designing filters. To reduce the passband return loss and increase insertion loss of the filter, the agent of the first task learned to gradually adjust the coupling coefficient between resonators at the current frequency, whose designed process was shown in Fig.3(a-c). The agents of the second task and the third task firstly learned to adjust the length of the resonators to achieve the goal of moving the center frequency, and then adjust the coupling coefficient between resonators to reduce the passband return loss and increase insertion loss, whose designed process were shown in Fig.3(d-i). To further verify whether the agent really learned to achieve the design target by adjusting the resonant frequency and coupling coefficient of the resonators, when the dimension is limited, the forth task was designed in this paper. Although the design central frequency of Task 3 and Task 4 were basically close, compared with Fig.3g and Fig.3j, the agent of Task 4 learned to increase the length of the resonator indirectly by widening the two ends of the resonators, thereby reducing the central frequency of the filter. It is worth mentioning that all the agents trained by this architecture converged in 10000 action steps and convergent evidence suggests that these agents discover relatively long-term strategies. In contrast, if only the traditional deep reinforcement learning model (see Methods for deep RL + Task 1) was used, there was still no sign of convergence after $10^6$ steps training, which showed the superiority and efficiency of the proposed architecture.

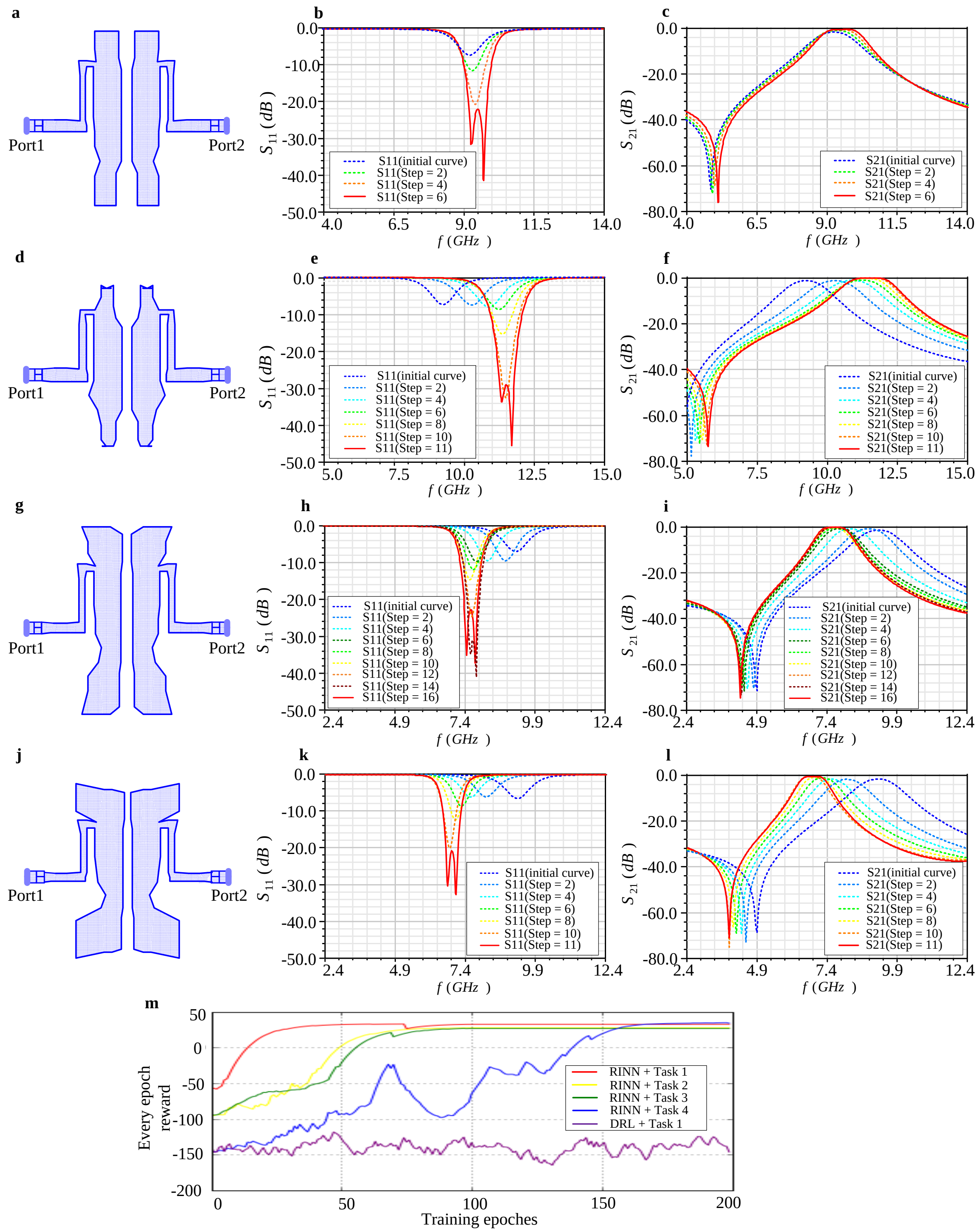

**Figure 3 | Filter design process based on RINN architecture**. **a-c**, An optimized filter model for the first task and its return loss (S11) and insertion loss (S21) change process separately. **d-f**, An optimized filter model for the second task and its return loss (S11) and insertion loss (S21) change process separately. **g-i**, An optimized filter model for the third task and its return loss (S11) and insertion loss (S21) change process separately. **j-l**, An optimized filter model for the forth task and its return loss (S11) and insertion loss (S21) change process separately. **m,** Learning curves of the four tasks. The learning speed of the agent was related to the complexity of the design task. The more complex the design task, the slower the learning speed of the agent.

## Antennas design based RINN

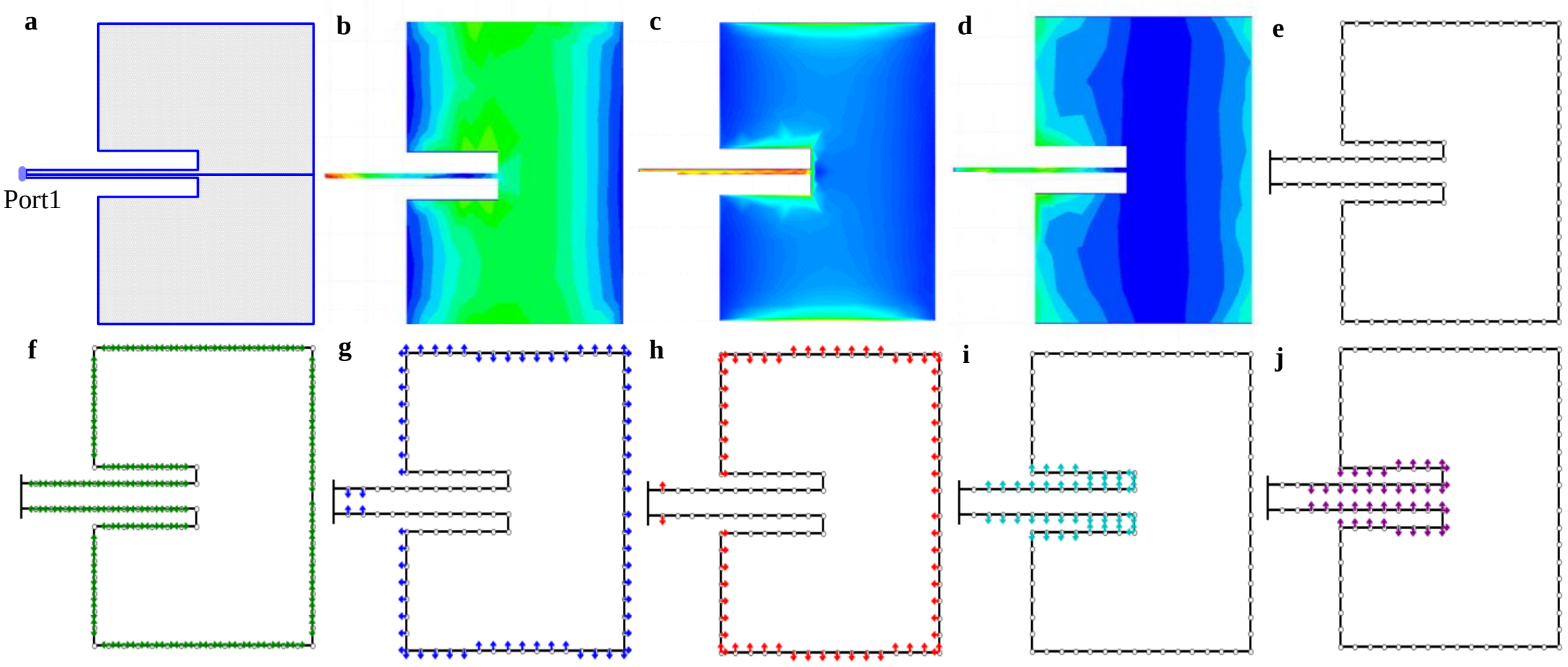

**Figure 4 | Clustering visualization results of antennas. a**, Antenna model. **b**, Surface current density distribution. **c**, Magnetic field distribution. **d**, Electric field distribution. **e**, Meshed model. **f-j**, Visualization results of mesh vertex clustering.

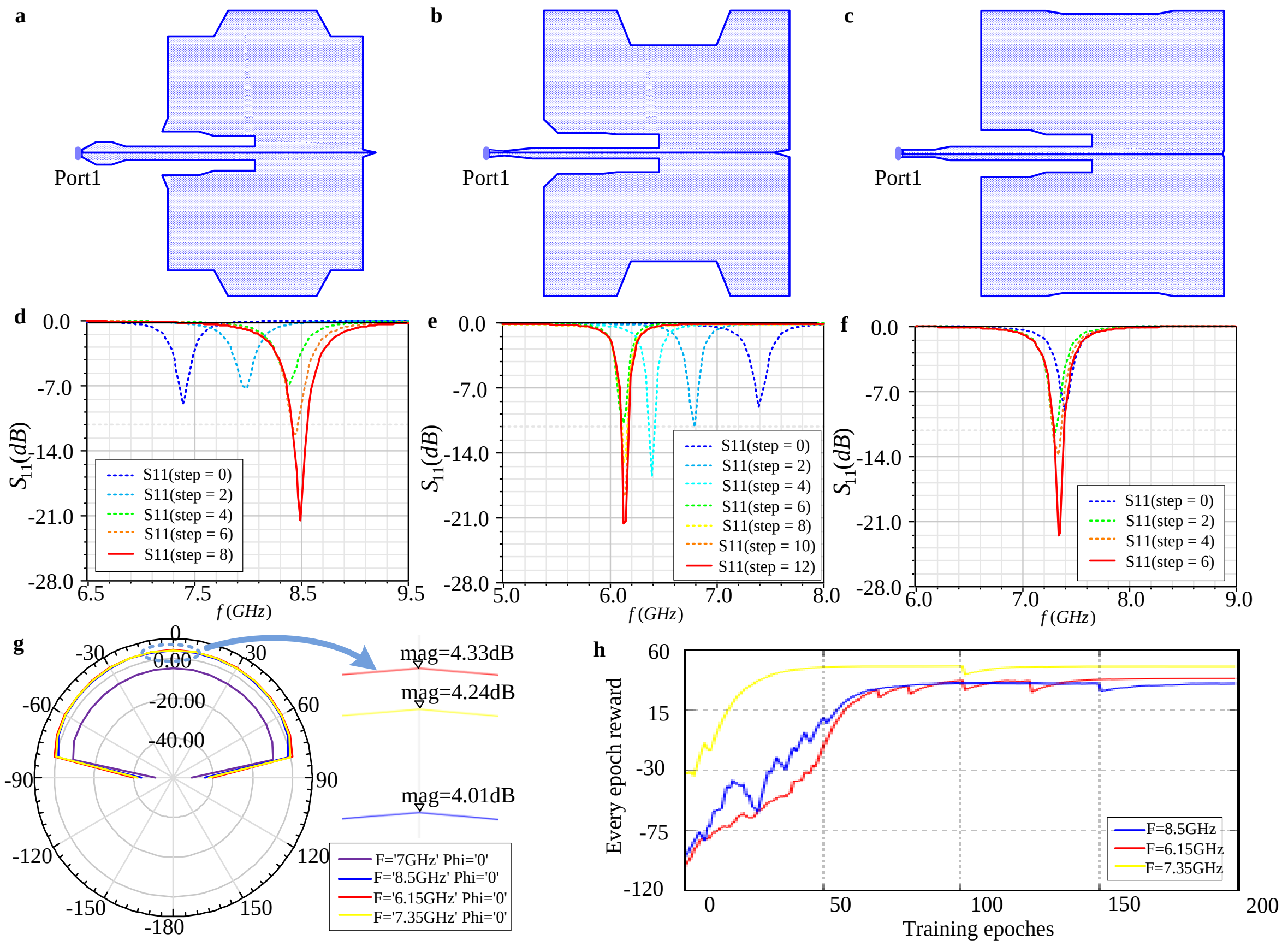

**Figure 5 | a-c**, Three antenna models, whose central frequencies are 8.5GHz, 6.15GHz and 7.35GHz respectively, designed by agents. **d-f**, Looking at the process of designing antennas by agents from the variation of return loss curves (S11). **g**, The gain pattern of all antennas. **h**, The learning curves of the three antenna models.

To further testify the generalization of RINN architecture, we tried to design antennas using it. The central frequencies of the antennas were 8.5 GHz, 6.15 GHz and 7.35 GHz, respectively. Meanwhile, the values of S11 at the central frequency were less than -20 dB and the gains were greater than 3.5 dB. The designed antenna is shown in Fig.4a: with side feed, the substrate is gallium arsenide material with a dielectric constant of 12.9 and gold for conductor. The clustering visualization results of the antenna are shown in Fig.4 (f-j), and the clustering method was the same as that of the filter. According to the theory of antenna (see Methods) and the distribution of electromagnetic field in Fig.4 (b-d), the key factors affecting the performance of antenna are radiation patch and feed line, whereas our proposed method also divides radiation patchesand feeders into two types separately.

As depicted in Fig.5, the agents trained by RINN successfully captured the main features of the antenna without any human knowledge and learned to perform a long series of correct actions in designing the antenna that succinctly express the causal relationships that give rise to their observations. The agents have successfully designed three antenna models with different frequencies based on the policy they have learned. From the process of designing the antenna, it could be seen that the radiation patch mainly affects the central frequency and the feeder mainly affects the input impedance. These coincided with the theory of rectangular patch antenna and the distribution of electromagnetic field. It further proved the scalability and interpretability of RINN.

## Compared to human engineers

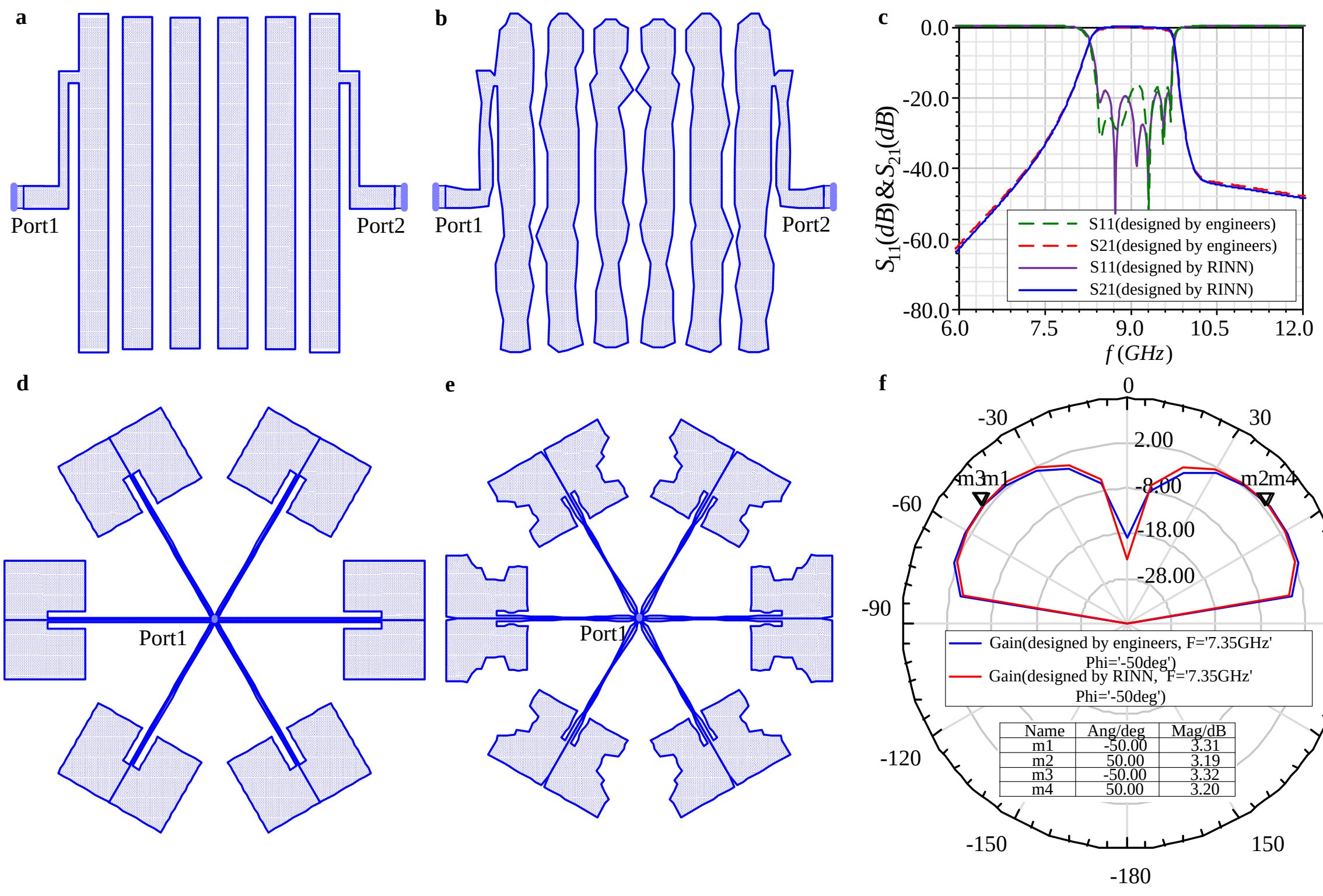

**Figure 6 | a**, A six-order filter model designed by engineers. **b**, A six-order filter model designed by RINN. **c**, the return loss curve (S11) and insertion loss curve (S21) of **a** and the return loss curve (S11) and insertion loss curve (S21) of **b**. **d**, A six-patch antenna model designed by engineers. **e**, A six-patch antenna model designed by RINN. **f**, The 7.35 GHz gain curves of **a** and **b**.

We compared RINN with a professional engineer by performing two experiments: first, a six-order filter whose reflection loss was less than -15dB and the insertion loss was greater than -1dB under the central frequency of 9.1GHz and the bandwidth of 1.2GHz; second, a six-patch antenna whose gain was greater than 3dB at a central frequency of 7.35 GHz. From the comparison (see Fig. 6) between the MWIC models designed by human engineers and those designed by RINN, we can see that the models designed by human engineers are regular and their number of parameters are limited. The model designed by RINN is irregular, with more parameters and higher degree of freedom, and its shape is more like that formed by nature. Indeed, the RINN agent is capable of learning to abstract the key parameters which affects the circuit performance and master a diverse array of design tasks. Consquently, the RINN agent, receiving only the meshed filter matrixes and S parameter matrixes as inputs, was able to achieve a level comparable to that of a professional engineer.

# Discussion

This paper illustrates a state-of-the-art machine learning techniques, which has the ability of induction and summarization, and can train agents to design microwave integrated circuit models completely independent of human engineers' experience. Besides, it can handle complex visual input while learning high-dimensional environmental information. The most distinct innovative aspect of the approach proposed herein includes that we integrate the clustering algorithm (induction of the relationship between the various parts of MWIC) and the reinforcement learning neural network (learning and designing MWIC) into a simple workflow. The workflow includes meshed shape and calculating the effect of each mesh change on the circuit performance. Clustering algorithm is designed to cluster these mesh changes, and reinforcement learning neural network is framed to train and understand them. In addition, as a theoretical verification, this architecture is conceived to design filters and antennas, and encouraging results are obtained. We find that the architecture can quickly reason about MWIC design steps and has good generalization. Moreover, the integrated circuit model designed with this architecture is interpretable, which is very helpful to engineers. This architecture can not only reveal the design methods, but also provide ideas for engineers to study the theory of MWIC.

The architecture proposed in this paper forms a closed loop process of end-to-end training in the stage of design, simulation and optimization of MWIC models. More importantly, the training process will not be affected by the mesh number of the model increasing or decreasing. For example, when the mesh number increases, we can judge the cluster of the new mesh by the trained clustering algorithm. Owing to our method can run on general mesh, it therefore has good scalability. In the future work, we plan to extend this method to the field of mechanical wave and mechanics, which will make it more universal.

# References

1. Larson L E. Radio frequency integrated circuit technology for low-power wireless communications[J]. IEEE Personal Communications,5(3):11-19 (1998).
2. Saxena A, Sahay B. Computer Aided Engineering Design[M]. Institution of Electrical Engineers, (1984).
3. A. M. Niknejad and R. G. Meyer, Design, Simulation and Applications of Inductors and Transformers for Si RF ICs. Dordrecht, The Netherlands: Kluwer Academic Publishers (2000).
4. S. K. Mandal, S. Sural, and A. Patra, ANN and PSO-based synthesis of on-chip spiral inductors for RF ICs, IEEE Trans. Comput. Aided Design Integr. Circuits Syst., vol. 27, no. 1, pp. 188–192, (2008).
5. S. Koziel, "Surrogate-based optimization of microwave structures using space mapping and kriging," in Proc. 39th Eur. Microw. Conf., Sep.–Oct. pp. 1062–1065, (2009)
6. A. Nieuwoudt and Y. Massoud, "Variability-aware multilevel integrated spiral inductor synthesis," IEEE Trans. Comput.-Aided Design Integr. Circuits Syst., vol. 25, no. 12, pp. 2613–2625, Dec. (2006).
7. S. Mandal, A. Goyel, and A. Gupta, "Swarm optimization based on-chip inductor optimization," in Proc. Int. Conf. Comput. Devices Commun, pp. 1-4, (2009).
8. J. W. Bandler, Q. S. Cheng, S. A. Dakroury, A. S. Mohamed, M. H. Bakr, K. Madsen, and J. Sondergaard, "Space mapping: The state of theart," IEEE Trans. Microw. Theory Tech., vol. 52, no. 1, pp. 337–361, (2004).
9. Ceperic V, Baric A. Modeling of analog circuits by using support vector regression machines[C]// IEEE International Conference on Electronics, Circuits and Systems. IEEE, 391-394, (2004).
10. S. Koziel and J. W. Bandler, "Space mapping with multiple coarse models for optimization of microwave

components," IEEE Microw.Wirel. Compon. Lett., vol. 18, no. 1, pp. 1–3, (2008).
11. A R Mirzai，C F N Cowan，T M Crawford． Intelligent alignmerit of． Waveguide falters using a machine learning approach． IEEE Transactions Microwave Theory and Technique, 37(1):166-173, (1989)．
12. Horng T S, Wang C C, Alexopoulos N G. Microstrip circuit design using neural networks[C]// Microwave Symposium Digest, 1993. IEEE MTT-S International. IEEE, vol.1., pp. 413-416, (1993).
13. Yang Y, Hu S M, Chen R S. A combination of FDTD and least ‑ squares support vector machines for analysis of microwave integrated circuits. Microwave & Optical Technology Letters, ,44(3): 296-299, (2005).
14. Güneş F, Tokan N T, Gürgen F. A knowledge-based support vector synthesis of the transmission lines for use in microwave integrated circuits. Expert Systems with Applications, 37(4):3302-3309, (2010).
15. Guney K, Sarikaya N. Concurrent Neuro-Fuzzy Systems for Resonant Frequency Computation of Rectangular, Circular, and Triangular Microstrip Antennas. 84:253-277, (2008).
16. J. Rayas-Sanchez, "EM-based optimization of microwave circuits using artificial neural networks: The state-of-the-art," IEEE Trans. Microw. Theory Tech., vol. 52, no. 1, pp. 420–435, (2004).
17. Stratigopoulos H G, Makris Y. Error Moderation in Low-Cost Machine-Learning-Based Analog/RF Testing[J]. IEEE Transactions on Computer-Aided Design of Integrated Circuits and Systems, 27(2):339-351, (2008).
18. Liu B, Zhao D, Reynaert P, et al. Synthesis of Integrated Passive Components for High-Frequency RF ICs Based on Evolutionary Computation and Machine Learning Techniques[J]. IEEE Transactions on Computer-Aided Design of Integrated Circuits and Systems, 30(10):1458-1468, (2011).
19. Park S J, Bae B, Kim J, et al. Application of Machine Learning for Optimization of 3-D Integrated Circuits and Systems[J]. IEEE Transactions on Very Large Scale Integration Systems,, 25(6):1856-1865, (2017).
20. Y. Gil, M. Greaves, J. Hendler, H. Hirsh, Amplify scientific discovery with artificial intelligence. Science 346, 171–172 (2014).
21. Musib M, Wang F, Tarselli M A, et al. Artificial intelligence in research.[J]. Science, 357(6346):28-30, (2017).
22. Mohaghegh S D. Reservoir simulation and modeling based on artificial intelligence and data mining (AI&DM)[J]. Journal of Natural Gas Science & Engineering, 3(6):697-705, (2011).
23. Caron, Mathilde, et al. Deep Clustering for Unsupervised Learning of Visual Features. https://arxiv.org /abs/1807.05520 (2018).
24. Collobert R, Weston J. A unified architecture for natural language processing:deep neural networks with multitask learning[C]// International Conference on Machine Learning. ACM, 2008:160-167.
25. M. Krakovsky, Reinforcement renaissance. Commun. ACM. 59, 12–14 (2016).
26. Mnih, V. et al. Human-level control through deep reinforcement learning. Nature, 518, 529-533 (2015).
27. K. De Asis, J. F. Hernandez-Garcia, G. Z. Holland, R. S. Sutton, Multi-step reinforcement learning: A unifying algorithm, http://arxiv.org/abs/1703.01327 (2017).
28. Mnih, V., Kavukcuoglu, K., Silver, D., Graves, A., Antonoglou, I., Wierstra, D., Riedmiller, M.A., Playing atari with deep reinforcement learning. CoRR abs/1312.5602, (2013).
29. D. Silver, A. Huang, C. J. Maddison, A. Guez, L. Sifre, G. van den Driessche, J. Schrittwieser, I. Antonoglou, V. Panneershelvam, M. Lanctot, S. Dieleman, D. Grewe, J. Nham, N. Kalchbrenner, I. Sutskever, T. Lillicrap, M. Leach, K. Kavukcuoglu, T. Graepel, D. Hassabis, Mastering the game of Go with deep neural networks and tree search. Nature 529,484–489 (2016).
30. Lijun Wu, Fei Tian, Tao Qin, Jianhuang Lai, Tie-Yan Liu. A Study of Reinforcement Learning for Neural Machine Translation. https:arxiv.org/abs/1808.08866. (2018).
31. Popova M, Isayev O, Tropsha A. Deep reinforcement learning for de novo drug design:[J]. Science Advances, 4(7), ( 2018).
32. Volodymyr Mnih, AdriÃ˘ a PuigdomÃ´ lnech Badia, Mehdi Mirza, Alex Graves, Timothy P. Lillicrap, Tim Harley, David Silver, and Koray Kavukcuoglu. Asynchronous methods for deep reinforcement learning. In Proc. of Int'l Conf. on Machine Learning, ICML, (2016).

# Method

### Meshed model

A meshed model is built by dividing the shape of MWIC, that is, through a series of vertices composed of curved surface or plane to describe the shape of the circuit. Extended Data Fig.1 is a four-vertex planar mesh. Four actions can be performed on the vertices. The specific changes are shown in Extended Data Fig.1(a-d).

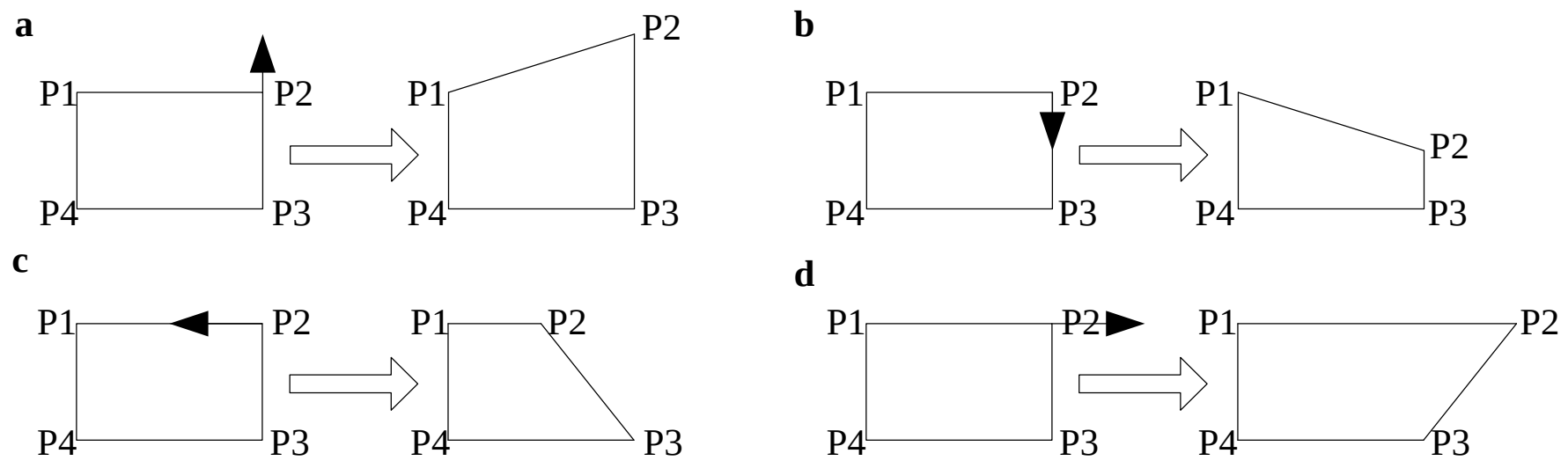

**Extended Data Figure 1 |** Schematic diagram of the change of planar mesh in four vertices

## Clustering algorithm

Along with the complexity of MWIC increasing, the optimization number of the MWIC about the shape and the size becomes larger. Supposing the meshed MWIC model has 500 vertices, and each vertex is executed four actions: upper, lower, left and right. The action space is 2000 (500*4), but the solution space is about 4^500. The complexity of the solution is far greater than that of Go. Therefore, the traditional reinforcement learning algorithm has to pay a heavy price for designing MWIC model. Clustering algorithm is a method to discover data distribution and existence patterns. It divides a set of objects into several subsets composed of similar objects according to attribute values. This unsupervised clustering process is widely applied in similar search, customer segmentation, pattern recognition, trend analysis and so on. In this paper, design actions are clustered by clustering algorithm which divides the change set of mesh vertices into several clusters.

There are many kinds of clustering algorithms, such as hierarchical, partition based and density based algorithms. For simplicity, in this paper, $k$-means algorithm based on partition was used to analyze the key factors of affecting MWIC performance. $k$-means takes n vectors $x_i$ ($i$=1,2,…,n) as input and clusters them into k distinct groups $G_i$ ($i$=1,2,…,$k$) based on a geometric criterionto minimize the objective function of non-similarity index. More precisely, when choosing the Euclidean distances between the vector $x_l$ from the group $G_i$ and the respective cluster center $c_i$ of $G_i$ as the similarity, the objective function can be defined as follows:

$$J = \sum_{i=1}^{k} J_i = \sum_{i=1,}^{k} \sum_{x_l \in G_i} \| x_l - c_i \|^2 \tag{1}$$

## Deep RL neural network model

The deep RL neural network model is addressed with the Asynchronous Advantage Actor-Critic (A3C) algorithm that relies on learning both a policy $\pi(a_t | s_t; \theta_p)$ and value function $V(s_t; \theta_v)$ given a state observation $s_t$. Both the policy and value function share all intermediate representations $\theta$, the loss function is as follows:

$$\nabla_{\theta_p} \log\pi(a_t | s_t; \theta_p) A(s_t, a_t; \theta, \theta_v) + \beta \nabla_{\theta_p} H(\pi(s_t; \theta_p)) \tag{2}$$

where $H$ is entropy and $\beta$ is a regularization coefficient of entropy. The entropy of policy $\pi$ is added to the loss function to encourage exploration and prevent the model from falling into local optimum. $A(s_t, a_t; \theta, \theta_v)$ is defined as an Advantage function:

$$A(s_t, a_t; \theta, \theta_v) = R_t - V(s_t; \theta_v) \tag{3}$$

where $R_t$ is a cumulative reward defined as:

$$R_t = \sum_{i=0}^{k-1} \gamma^i r_{t+i} + \gamma^k V(s_{t+k}; \theta_v) \tag{4}$$

The complete deep RL network architecture is as follows. The shared network is composed of F1 and F2, from which we predict the policy and value function. The F1 network is an encoder model with 2 convolutional layers followed by a fully connected layer. The input observation (Meshed filter matrixes) of F1 is first processed through two convolutional layers with 8 and 16 kernels, $3 \times 3$ kernel sizes and a stride of 1, followed by a rectified linear unit (ReLU) activation function. The fully connected layer has 64 hidden units. The F2 network is an encoder model with 2 MLP layers with 512 and 256 hidden units, each followed by a tanh activation function, whose input observation is S parameter matrixes. The outputs of the policy module , the actor, is a 2 fully connected layers, each followed by a tanh activation function and a softmax activation function. The outputs of the value module , the critic, is a 2 fully connected layers, each followed by a tanh activation function and a single linear unit. The sizes of policy logits $\pi(a_t \mid s_t;\theta_p)$ and baseline function $V(s_t;\theta_v)$ are the number of action clusters and one separately. The policy logits were normalized and used as multinomial distribution from which the action is sampled.

**Reward of deep RL model**

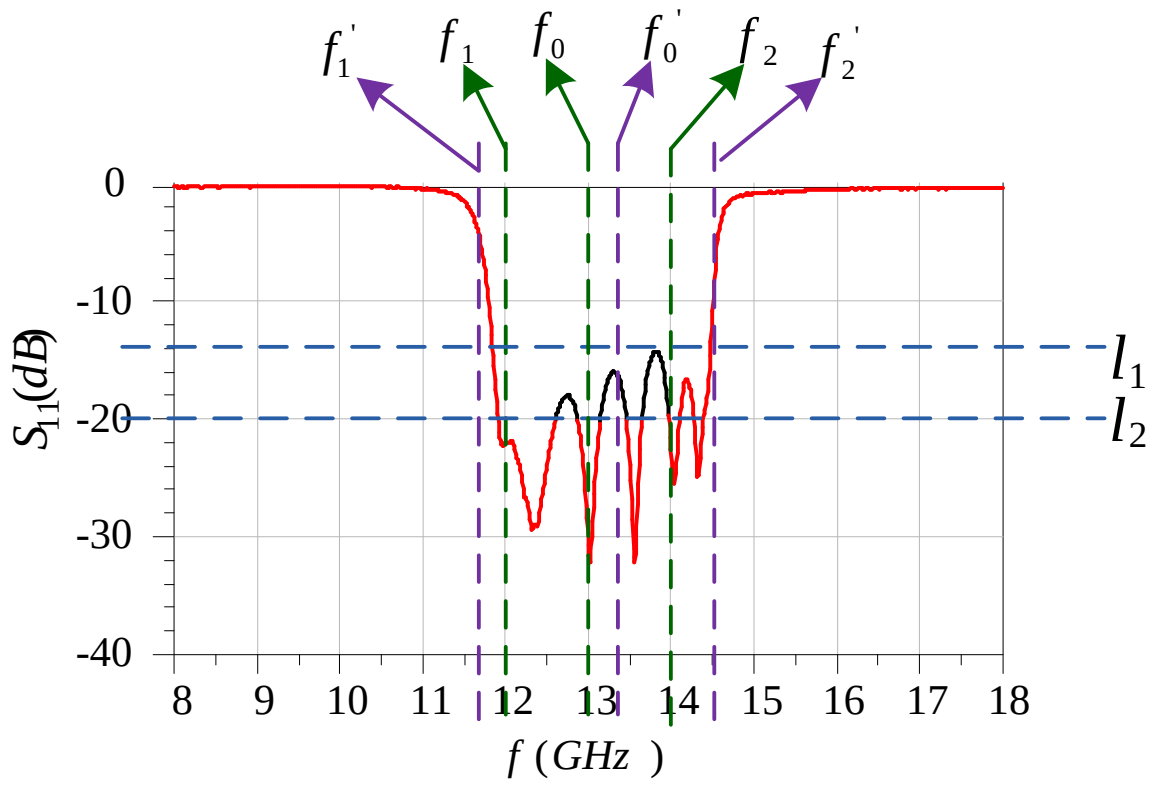

**Extended Data Figure 2 |** Reward function settings

The design parameters of MWIC are generally related to the central frequency $f_0$, passband frequency range $f_1 \sim f_2$, return loss and other electrical characteristics. So the reward of deep RL model is a function of the predicted property of the design parameters. First, the reward function related to the design parameters is calculated by the return loss S11. As shown in Extended Data Fig.2, when the center frequency $f_0'$ moves to the design parameters $f_0$, that is to say, when the distance between $f_0'$ and $f_0$ is getting smaller and smaller, we should give a positive reward. At the same time, the passband frequency range $f_1'$ and $f_2'$ are approaching to design targets $f_1$ and $f_2$ respectively, we also give positive rewards. The return loss (return loss is designed to be -20dB) of passband frequency range $f_1 \sim f_2$ is a curve between line $l_1$ (horizontal line at the maximum peak of the curve) and line $l_2$ (horizontal line at -20dB), that is, the black curve in Extended Data Fig.2 is represented by return loss(Echo Loss, EL). The smaller the mean, the greater the positive reward. Finally, the reward function is obtained by integrating the above reward into the following reward function:

$$r = \frac{1}{|f_0 - f_0'|} * \beta_1 + \frac{1}{|f_1 - f_1'| + |f_2 - f_2'|} * \beta_2 + \left( \frac{1}{N} \sum_{i=1}^{i=N} EL_i \right) * \beta_3 \tag{5}$$

**Research of microwave transmission line based on RINN**

Microwave transmission line is a transmission system for transmitting microwave information and energy. It has the function of guiding electromagnetic wave to transmit in a certain direction. Transmission line is the basis of MWIC, so this part uses the clustering algorithm to study how the size of each part of the transmission line affects the performance.

The ratio of voltage to current at any point on a transmission line is called its impedance at that point, but the microwave impedance can not be measured directly which can only be measured indirectly by means of input impedance and Voltage Standing Wave Ratio. The S parameter is a network parameter based on the relation between incident wave and reflected wave. It is used to describe the N-port Microwave network. S parameters are used to study the performance of transmission lines. The transmission line can be equivalent to a two-port device, so the S parameter matrix is interpreted based on the two-port network. Extended Data Fig.3 shows the S parameter matrix of the two port network, where S11 (S11=b1/a1= Reflected power / incident power) indicates the input reflection coefficient usually termed the return loss and S21(S21=b2/a1 = Output power / input power) indicates the forward transmission gain coefficient commonly referred to as insertion loss. Therefore, the RINN architecture realizes the MWIC design using S parameters. Symmetrical transmission lines have the same characteristics as S11 and S22, S21 and S12. Finally, S11 and S21 are used to study the performance of transmission lines.

Extended Data Fig.4a is a transmission line model whose meshed model as shown Extended Data Fig.4b. Then the differential S parameter matrix was S0 minus S, where S0 was original S parameter matrix without actions and the S1 was the S parameter matrix by moving 0.12mm at each point according to the top and bottom. Finally, we cluster differential S parameter matrices and the result is shown in Extended Data Fig.5. From Extended Data Fig.5, we can see that the clustering algorithm clearly divides the model into 4 typical categories according to its length and width (excluding the action cluster that has little effect i.e. Extended Data Fig.5a). The length mainly affects the frequency variation, and the width mainly affects the characteristic impedance.

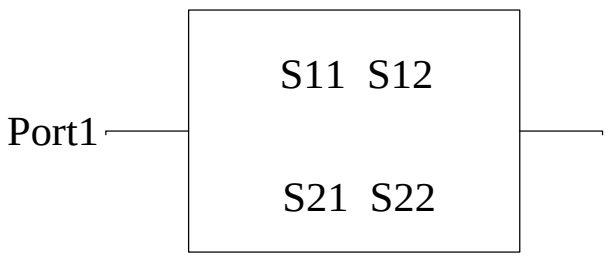

**Extended Data Figure 3 |** S parameter matrix

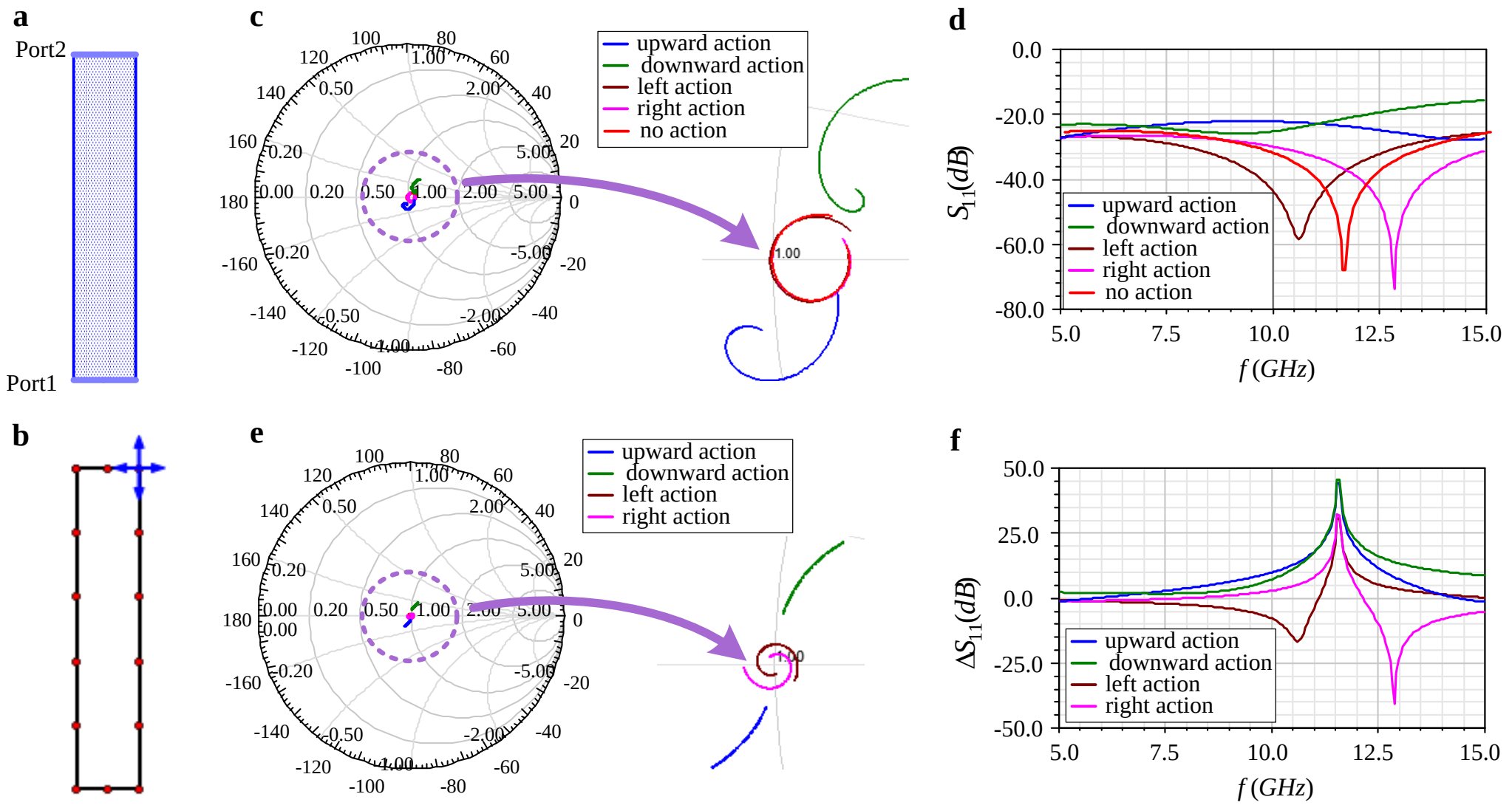

**Extended Data Figure 4 | a**, Transmission line model. **b**, Meshed model. **c**, Obtained Smith Chart of S11 by performing actions according to the arrow of Extended Data Fig.4b. **d**, Obtained Curve of S11 by performing actions according to the arrow of Extended Data Fig.4b. **e**, Differential Smith Charts which are the difference between the original Smith Charts without actions and the Smith Charts by performing actions. **f**, Differential S11 curves which are the difference between the original S11 curve without actions and the S11 curve by performing actions.

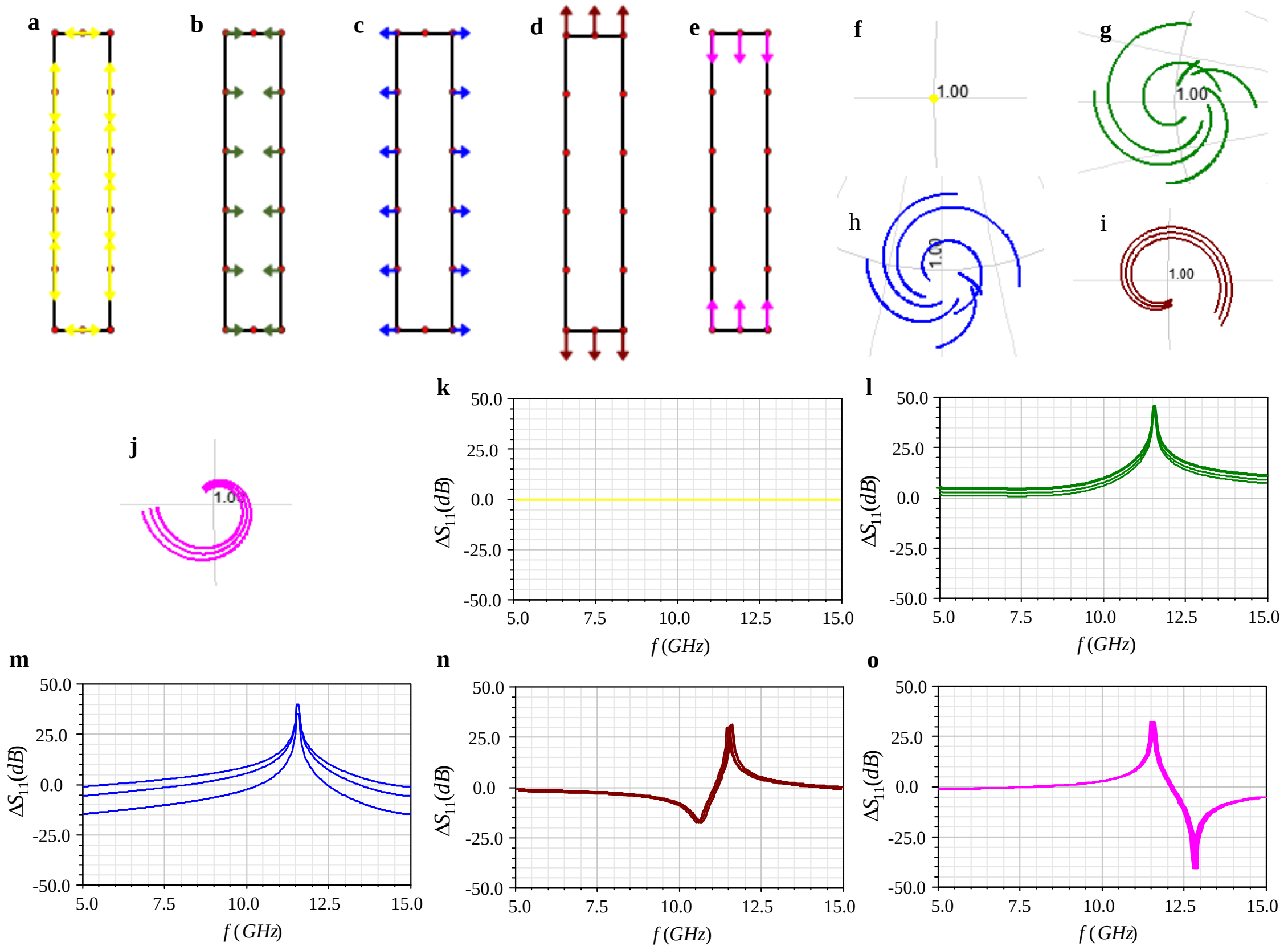

Extended Data Figure 5 | Clustering results. **a-e**, Visualization results of typical action clusters on the transmission line. **f-j**, Differential Smith Charts of the typical action clusters. **k-o**, Differential S11 curves of the typical action clusters.

## Theory of filters

The filter schematic of Fig.2a is shown in Extended Data Fig.6, which includes Source (S), Load (L), Resonator 1 (R1), Resonator 2 (R2). The coupling coefficient between S and R1 is M0, between R1 and R2 is M1, and between R2 and L is M2. Only when the coupling coefficients M0, M1 and M2 satisfy a certain proportion, the filter can run with reasonably good performance. Besides, the length of the resonators also affect the performance of the filter, that is, the resonance frequency of the filter increases or decreases when R1 and R2 increase or decrease simultaneously. The filter in Fig.2a is symmetrical, so M0 = M2, R1 = R2. According to the filter theory, there are four key factors affecting the filter: resonant frequency increases, resonant frequency decreases, coupling ratio increases and coupling ratio decreases. From the design process of the agents, we can see that these four factors were adjusted when the tasks were gradually up to the standard, which verified the inductivity and learnability of the proposed architecture. For example, from the surface current density distribution of Fig.2b, it is known that the coupling signal of the filter is mainly concentrated at the middle of the resonators, whereas the agent can reduce the distance between the two resonators by doing the action of Fig.2i, that is, increasing the coupling coefficient between the two resonators.

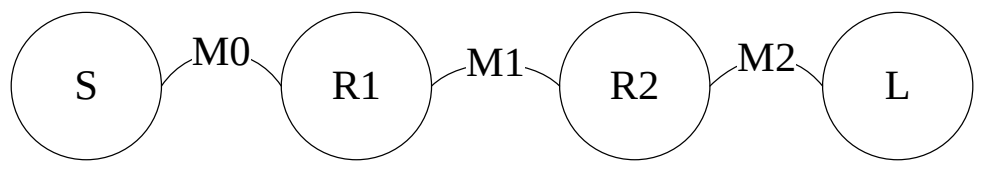

**Extended Data Figure 6** | Bandpass filter schematic diagram

## Deep RL algorithm to filterdesign

In order to verify the efficiency of the RINN, we train the meshed filter shown in Fig.5e only with deep reinforcement learning algorithm. The deep reinforcement learning neural network structure is the same as that of the deep RL neural network modelof the RINN, but the only difference is the output of the policy network whose number of actions depends on the number of vertices in the meshed filter.

**Antenna theory**

The performance of the antenna is related to location of the feed line, geometry and size. Extended Data Fig.7a is the structure of the antenna. The length of the radiation patch is $L$, the width is $W$, the thickness of the dielectric substrate is $h$, and the working wavelength is $\lambda$. If radiation patches, dielectric substrates and grounding boards are regarded as a low impedance transmission line with a length of $0.5\lambda$, the open circuit is formed at both ends of the radiation patch. The electric field at the two open ends of the radiation patch can be decomposed into the horizontal component (Extended Data Fig.7b) and the vertical component (Extended Data Fig.7c) with respect to the grounding plate only considering the main mode excitation. The horizontal component can be equivalent to two gaps in a perfectly conducting plane and the gaps' width are $\Delta L$, so the antenna can radiate electromagnetism energy along the direction of width $W$ at gaps.

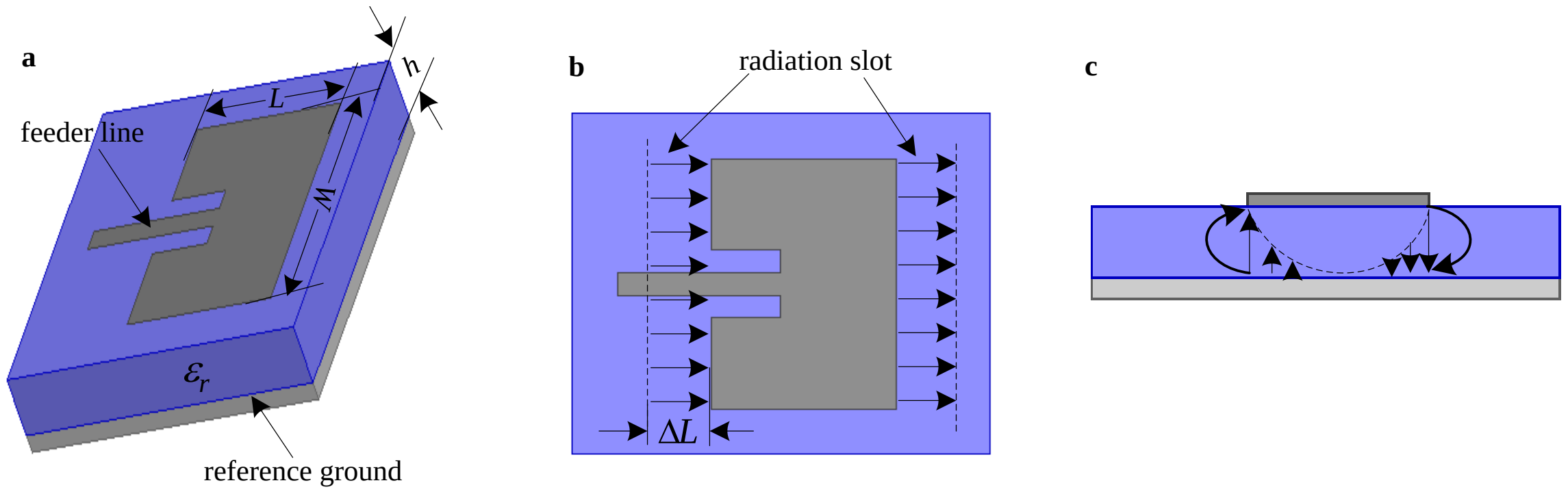

**Extended Data Figure 7 | a**, Antenna structure diagram. **b**, Antenna top view. **c**, Antenna side view.